\documentstyle[titlepage,12pt]{article}
\newcommand{\be}{\begin{equation}}
\newcommand{\ee}{\end{equation}}
\newcommand{\bear}{\begin{eqnarray}}
\newcommand{\ear}{\end{eqnarray}}

\newcommand{\bfsp}{{\mbox{\boldmath$\sigma_+$}}}
\newcommand{\bfsm}{{\mbox{\boldmath$\sigma_-$}}}

\begin{document}
\begin{titlepage}
hep-ph/9608364 \hfill
PITHA 96/26 \rightline{August 1996} \rightline{ }

\vspace{0.8cm}
\begin{center}
{\bf\LARGE CP Violation Beyond the Standard Model\\}
{\bf\LARGE  and Tau Pair Production in
$e^+ e^-$ Collisions
 \\}
\vspace{2cm}

\centerline{ \bf W. Bernreuther$^{a}$\footnote{Research supported  by BMBF
contract 057AC9EP.},
A. Brandenburg$^a$\footnote{Research supported  by Deutsche
Forschungsgemeinschaft.},
and P. Overmann$^b$}
\vspace{1cm}
\centerline{$^a$Institut f. Theoretische Physik,
RWTH Aachen, D-52056 Aachen, Germany}
\centerline{$^b$Institut f. Theoretische Physik,
Universit\"at Heidelberg, D-69120 Heidelberg, Germany}
\vspace{3cm}

{\bf Abstract:}\\
\parbox[t]{\textwidth}
{We show that the
CP-violating dipole form factors of the tau lepton can be of
the order of $\alpha/\pi$ in units of  the length scale set by
the inverse $Z$ boson mass. We propose a few observables which
are sensitive to these form factors at LEP2  and higher
$\rm{e^+e^-}$ collision energies.}

\end{center}
\end{titlepage}
\newpage

\section{Introduction}
In order to clarify whether or not the Kobayashi-Maskawa phase \cite{Kob}
in the quark mixing matrix
is the sole cause of CP violation in nature, as many CP tests as possible
-- which are in particular
sensitive to other conceivable CP-violating interactions -- should be
performed, also outside of the
kaon system.  One possibility is the search for leptonic CP violation.
As to the tau lepton a  number of proposals have been made in this connection
\cite{Stod}-\cite{Tsai}.
The OPAL \cite {O1,O2} and ALEPH \cite{A1,A2,A3} detector groups  have
demonstrated by detailed
investigations of $Z\to\tau^+\tau^-$  at LEP that sensitive CP symmetry tests
at the few per mill level can be performed in high energetic $\rm{e^+e^-}$
collisions.
Specifically they have obtained upper bounds on the CP-violating weak
dipole form
factor of the tau lepton that have recently reached \cite{O2,A3,ST} a level
well below $10^{-17}e$ cm.

In this letter we investigate for a number of CP-violating interactions
the possible size of the CP-violating dipole form factors of the tau
lepton. Moreover we propose
a few observables that are sensitive to these form factors at LEP2 energies
and at energies of a presently discussed high luminosity linear
$\rm{e^+e^-}$ collider.
\section{Models}

\noindent
Detectable CP violation in tau production and decay
requires new CP--violating interactions involving leptons.
Here we discuss only possible effects in tau pair production. (For a
discussion of tau decay, see
\cite{GNelson, Nelson, Hagiwara, Tsai}.)
These interactions  would induce in   the $e^+ e^-\to \tau^+\tau^-$
scattering amplitude
electric (EDM) and weak (WDM) dipole form factors of the tau lepton through
radiative
corrections. For a number of models
the CP-violating contribution to the one-loop ${\cal T}$ matrix element
is, in the limit of vanishing electron mass,  of the form

\begin{equation}
{\cal T}_{CP} = -e \left[
J^\mu_\gamma    {d_\tau^\gamma(s) \over s}
+ {1 \over s_W c_W }J^\mu_Z
{d_\tau^Z (s) \over s-m_Z^2 }\right]
\times \bar u_\tau(k_\tau) \sigma_{\mu\nu}
\gamma_5 k^\nu v_\tau(k_{\bar\tau}) ,
\label{TCP}
\end{equation}

\noindent
where $J_\gamma^\mu = - \bar v_e \gamma^\mu u_e,$
$J_Z^\mu = -\bar v_e \gamma^\mu (1-\gamma_5) u_e/4-s^2_W J_\gamma^\mu,$
$s_W = \sin\theta_W, c_W = \cos\theta_W$,
$k=k_\tau + k_{\bar\tau},$ $s=k^2$. (In the vicinity of the
$Z$ resonance the $Z$ width must of course be taken into
account in the $Z$ propagator.)
The form factors are ultraviolet finite if the interactions are renormalizable.
Depending on the model and on the c.m. energy  the $d_\tau^{\gamma,Z}(s)$
can have also absorptive parts.

Since it is known experimentally
that quarks and leptons are pointlike particles
up to a scale of order $10^{-16}$cm
one may consider the length scale set by the inverse $Z$ boson mass to be
the natural scale
for quark and lepton EDMs and WDMs.  Therefore we write
\begin{equation}
d_\tau  = e \frac{\delta}{m_Z}.
\label{delta}
\end{equation}
In the models discussed below $d_\tau^{\gamma,Z}$
are generated by radiative corrections at one loop. In general they may be
expected to
be of the order of a typical electroweak correction, that is, of order
$\alpha/\pi$.
Moreover $d_\tau^{\gamma,Z}$ are chirality-flipping form factors which are,
on general
grounds, proportional to some fermion mass $m_F$.  Thus we have schematically
$\delta \sim (\alpha/\pi)\times(m_F/m_Z)$. The fermion mass need not be the
tau mass but can be
the mass of a fermion $F\neq\tau$ in the loop, which may be much larger
than $m_{\tau}$. (In the case of the one-loop
contributions to the EDM $m_F$ is always the mass of the fermion
 in the loop.) Hence there is no
a priori argument that $\delta$ must be suppressed by powers of $m_{\tau}$ --
contrary to the claim of \cite{Tsai}.

\noindent These chirality flipping form factors lead to an incoherent
contribution
to   $d\sigma/d\cos\theta_\tau,$ which is proportional to
$\sin^2\theta_\tau$ and is bilinear in $d_\tau^{\gamma,Z}.$
For \\ $|\delta|\ll 1$ this distribution is
therefore not very sensitive. Moreover, it does not
constitute a CP test: a magnetic moment form factor  induces a term
$\sim \sin^2\theta_\tau$, too.
Obviously searches for a term (\ref{TCP}) in the scattering amplitude
should be done with
CP--odd observables  whose expectation values are (to good approximation)
linear
in $d_\tau^{\gamma,Z}.$

The extremely tiny upper bound on the electric dipole moment of the
electron \cite{Commins}
 may at first sight
discourage searches for CP violation in tau production.  However, these
searches make sense
because CP-violating interactions of non-universal strength are conceivable
that induce
a tau EDM and WDM being much larger than those of the electron.
A prototype of such an interaction is CP violation by an extended Higgs
sector, where the symmetry
breaking interactions are unrelated to the mixing of fermion generations
\cite{Lee}.
For two-Higgs doublet extensions of the Standard Model with natural flavour
conservation the
real and imaginary parts of the EDM and WDM were computed for the top quark
in \cite{BSP}.
The formulae given there can be readily transcribed to the tau lepton. We
get that in this type of models
$\delta$ may become as large as $10^{-4}$.

In supersymmetric extensions of the Standard Model the
$\tau$-$\tilde\tau$-neutralino couplings may contain  CP
phases and thereby generate a non-zero tau EDM and WDM at one loop.
($\tilde\tau$
denotes a scalar tau.) The chirality flip is provided by the
neutralino mass.  Applying the formulae of \cite{BO} we obtain
 that not too far away from the $\tilde\tau$ threshold
$\delta$ can be as large as a few$\times 10^{-4}$ in the case of the EDM.

Larger effects may be induced by leptoquarks.
Leptoquark bosons, which
mediate quark-lepton transitions,
 appear naturally in unified and composite models (see e.g. \cite{lepto}).
Here we are interested only
in  spin zero leptoquarks with  $SU(3)_C\times SU(2)_L\times U(1)_Y$ invariant
couplings to quarks and leptons, which are, moreover,
 baryon and lepton number conserving.  We consider two different types
of spin zero leptoquarks: a weak isodoublet $\chi=(\chi_1,\chi_2)$ with
quantum numbers $\chi(3,2,{7\over 6})$ (model I) and a weak isosinglet
$\chi_0$ with quantum numbers $\chi_0(3,1,-{1\over 3})$ (model II).
The corresponding interaction Lagrangians involving the fermions of the
third generation  are \cite{BRW}:
\begin{equation}
 {\cal L}_I = \lambda_1 ({\bar q}_L\cdot\chi)\tau_R + {\tilde\lambda}_1
(\chi^Ti\sigma_2 \ell_L) {\bar t}_R + h. c.,
\label{lepto1}
\end{equation}
\begin{equation}
 {\cal L}_{II} = \lambda_2 ({\bar t^c}_R\tau_R)\chi_0^{\dagger} +
{\tilde\lambda}_2 ({\bar q^c}_Li\sigma_2 \ell_L)\chi_0^{\dagger}
   + h. c.
\label{lepto2}
\end{equation}
\noindent Here $q_L=(t,b)_L$ and $\ell_L=(\nu_\tau,\tau)_L$, the label $c$
denotes charge conjugation, and $\sigma_2$ is the Pauli matrix acting in the
weak isospace.

While leptoquarks that couple to the first and second generation
of quarks and leptons
are strongly constrained in their masses and couplings (see, e.g.
\cite{DBC,H1}),
the bounds on third generation leptoquarks are less restrictive
\cite{H1,Ellis,Eboli}.
An analysis of radiative corrections to observables for $Z$ boson physics leads
to the conclusion that the masses of the doublet $\chi$
(which we assume to be degenerate in mass) and of $\chi_0$ cannot be
smaller than
about 200 GeV if the couplings of these bosons to $t$ quarks and $\tau$ leptons
are of weak interaction strength \cite{Ellis,Eboli}.

If ${\rm Im}({\tilde\lambda}_i^*\lambda_i)\neq$0 then the
$\tau t\chi_1$ and $\tau t\chi_0$ couplings in (\ref{lepto1}) and
(\ref{lepto2}),
respectively, are CP-violating\footnote{For discussions of other
CP-violating effects due to scalar leptoquarks, see for instance
\cite{Ber,CPlep}.}.
In this case the following EDM and WDM form
factors of the $\tau$ lepton are induced to one-loop aproximation:

\begin{eqnarray}
d^{\gamma}_\tau = e m_tN_C
\frac{{\mbox{Im}}({\tilde\lambda}_i^*\lambda_i)}{8\pi^2}\frac{1}
{s\beta_{\tau}^2}
\left[Q_t H(s) - Q_{\chi}K(s)\right],
\label{Glep}
\end{eqnarray}

\begin{eqnarray}
d^Z_\tau = \frac{e m_t N_C}{s_W c_W}
\frac{{\mbox{Im}}({\tilde\lambda}_i^*\lambda_i)}{8\pi^2}\frac{1}
{s\beta_{\tau}^2}
\left[(g_t^V H(s)- g_{\chi}
K(s)\right],
\label{Zlep}
\end{eqnarray}

\noindent with

\begin{eqnarray}
H(s)=B_0(s,m_t^2,m_t^2)-B_0(m_{\tau}^2,m_t^2,m_\chi^2)-\nonumber \\
(m_t^2-m_\chi^2-m_{\tau}^2)
C_0(s,m_t^2,m_\chi^2,m_t^2),\nonumber\\
K(s)=B_0(s,m_\chi^2,m_\chi^2)-B_0(m_{\tau}^2,m_t^2,m_\chi^2)- \nonumber\\
(m_{\tau}^2+m_\chi^2-m_t^2-s/2)
C_0(s,m_\chi^2,m_t^2,m_\chi^2).
\label{K}
\end{eqnarray}

\noindent In (\ref{K}) $B_0$ and $C_0$ denote the standard scalar
2- and 3-point functions \cite{tHV}. Further, $\beta_\tau =
(1-4m^2_\tau/s)^{1/2}$,
 $N_C$=3, $Q_t=2/3$, $g_t^V=1/4-2s^2_W/3$, and
$g_{\chi}=T_3^{\chi}-Q_{\chi} s^2_W$.
 The results for
models I and II are obtained by inserting into (\ref{Glep}),(\ref{Zlep})
the quantum numbers $(Q_\chi,T_3^\chi)=({5\over 3},{1\over 2})$ and
$(-{1\over 3},0),$ of $\chi_1$ and $\chi_0$, respectively.

The chirality flip is provided by the
mass of the top quark. The form factors (\ref{Glep}),
(\ref{Zlep}) cross, as functions of the c.m. energy, the $\chi\bar{\chi}$
and $t\bar{t}$ thresholds. Above the lower of the two the EDM and WDM develop
imaginary parts.
In order to illustrate the possible size
 of the form factors the real and imaginary parts of the EDM and WDM   are
plotted
in Fig.1 for the leptoquark doublet model with $m_t$= 180 GeV and choosing
$m_{\chi_1}$ =  200 GeV.
Using the results of  \cite{Ellis,Eboli} and taking the CP phase to be maximal,
we get
$|{\rm Im}({\tilde\lambda}_1^*\lambda_1)|\leq$ 0.44. Larger couplings
are tolerable if $\chi_1$ is heavier; but this would not increase the EDM
and WDM
as compared to the case exhibited in  Fig. 1.
>From a numerical analysis we conclude that
in the leptoquark doublet model
the real part of the EDM may be as large as $0.3\times 10^{-18}e$ cm above
the $Z$
resonance.
The real part of the WDM
is smaller than $\rm{{\mbox Re}d_{\tau}^{\gamma}}$ by a factor of about $4$.
The numerical value of
 $\rm{{\mbox Re}d_{\tau}^{\gamma}}$ in model II is smaller by a factor of
2.8 than in model I.

One may expect that the couplings of the scalar leptoquarks in
(\ref{lepto1}) and (\ref{lepto2}) are of the Higgs boson type.
Then the couplings of $\chi$ and $\chi_0$ would be proportional to the mass
of the
right-handed fermion involved. That is, $\lambda \sim m_{\tau}/M$ and
${\tilde\lambda} \sim
m_t/M$, where $M$ is some mass scale. Analogous relations hold for the
couplings to the first and to the second generation of quarks andleptons. Furthermore one may
expect that inter-generation couplings are suppressed by small mixing
angles and hence cannot
become more important than generation-diagonal couplings. If this is the
case, one
 gets the following scaling relation for the electron, muon, and tau dipole
moments:
\begin{equation}
d_e : d_{\mu} : d_{\tau} =  m_u^2 m_e : m_c^2 m_{\mu} : m_t^2 m_{\tau}
\label{scale}
\end{equation}
\noindent This relation indicates that the
tau dipole moments can be  of the order of a small electroweak radiative
correction
 -- i.e.  $|\delta| \simeq 0.001$ as obtained above --
whereas the electron EDM is severely suppressed by small fermion masses
and hence well below the experimental upper bound \cite{Commins} of
$4\times 10^{-27}e$ cm.
The above type of leptoquark couplings have another amusing feature. They
generate also EDMs and WDMs of $u$, $c$, and $t$ quarks which are smaller
in magnitude than the corresponding
moments of the charged leptons within the same generation.

As a final example we mention tau dipole moments due to  heavy Majorana
neutrinos.
These particles appear naturally, for instance, in  grand unified theories.
These
models are in addition endowed with an extended
Higgs sector.
If there are charged Higgs boson ($H^+$) couplings to a heavy neutrino
$N_{\tau}$ and the tau lepton,
\begin{equation}
 {\cal L}_N = (2\sqrt 2 G_F)^{1/2} (\beta_1 m_{\tau}
{\bar N}_{\tau}\tau_R + \beta_2 m_{N_{\tau}} {\bar N}_{\tau}\tau_L)  H^+  +
h. c.,
\label{neutr}
\end{equation}
with $\rm{{\mbox Im}(\beta_1\beta_2^*)}\neq 0$, then non-zero tau EDM and
WDM are generated.
The chirality flip comes from the mass of the $N_{\tau}$, which may be of
the order of a few hundred GeV.
The computation of the moments is straightforward.  Their maximal size
is of similar order of magnitude as in the leptoquark model above.

The above discussion shows that
the tau EDM can be of the order of
$d_{\tau}^{\gamma} = e \delta/m_Z$ with $|\delta| \simeq$ a few$\times
10^{-3}$. In the leptoquark models
the WDM is smaller by a factor of about four.

\section{CP-odd correlations}

\def\k{{\bf\hat k}}
\def\e{{\bf\hat e}}
\def\p{{\bf\hat p_+}}
\def\m{{\bf\hat p_-}}

\noindent
The above form factors can be traced with appropriate observables in tau
pair production.
Here we consider
unpolarized $e^+e^-$ collisions above the $Z$ boson resonance  and
 decays of $\tau^{\pm}$ into the following channels:

\begin{eqnarray}
e^+ + e^-
\to \tau^+ + \tau^-
\to A + {\bar B} +X,
\label{tau}
\end{eqnarray}
where  $A,B=\pi,\rho$, and $\ell = e, \mu$.
Generic CP symmetry tests for these reactions are as follows. Consider
observables ${\cal O}$ which change sign under a CP transformation.
One can prove  that in the case of unpolarized and transversely polarized
$\rm{e^+e^-}$
collisions and CP-invariant phase space cuts
\begin{eqnarray}
<{\cal O}>_{A \bar B} + <{\cal O}>_{B \bar A}
& \neq & 0
\label{CP}
\end{eqnarray}
is an unambiguous  signal of CP violation \cite{BeNa}.
More specifically,  non-zero
tau EDM and WDM induce a number of CP-odd  spin-momentum correlations in the
$\tau^+\tau^-$ system \cite{BBNO}.  For instance they lead to non-zero
expectation values
of the the following
CP- and T-odd observables involving the $\tau^{\pm}$
spins ($\sigma_{i\pm}$ are the Pauli matrices with
$\pm$ refering to the respective spin spaces, and
$\e$, $\k$ are  the directions of the incoming positron and of the $\tau^+$
in the overall
c. m. frame, respectively):
\begin{eqnarray}
{\cal O}_1 & = &
(\e\times \hat{\bf k})\cdot({\bfsp}-{\bfsm}),\nonumber \\
{\cal O}_2 & = & \hat{\bf k}\cdot {\bfsp}
(\e\times \hat{\bf k})\cdot{\bfsm}-
\hat{\bf k}\cdot {\bfsm}
(\e\times \hat{\bf k})\cdot{\bfsp}
\label{sigma}
\end{eqnarray}
Non-zero  $\rm{\mbox{Re}d^{\gamma,Z}_{\tau}}$
generate for instance a tau polarization normal to the scattering plane
which differs in sign for $\tau^+$ and $\tau^-$. This makes $<{\cal O}_1>
\neq 0$.
 Absorptive parts from CP-invariant interactions
in the scattering amplitude lead to  equal $\tau^{\pm}$ normal
polarizations and thus cancel in $<{\cal O}_1>$.
An analogous statement applies to the longitudinal-normal
spin-spin correlation $<{\cal O}_2>$.  A closer inspection
reveals that ${\cal O}_1$ has a higher sensitivity
to  $\rm{\mbox{Re}d^{Z}_{\tau}}$ than  to
$\rm{\mbox{Re}d^{\gamma}_{\tau}}$. For
${\cal O}_2$ the opposite holds.
If one takes the sums instead  of the
differences (\ref{sigma}) one projects onto CP-invariant absorptive parts.

The tau spins are analysed by the
decay distributions of the charged prongs.
Below we consider only the channels  $\pi\pi, \pi\rho,
\pi\ell, \rho\rho$, and $\ell\ell$ that have a good
$\tau$-spin analyzer quality.
Spin-momentum correlations like (\ref{sigma}) can be translated into
correlations among the momenta of the charged particles $A, \bar B$ and
the charge conjugated modes.

In \cite{BNO} a number of correlations involving momenta in the overall c.
m. frame
were computed for various $\rm{e^+e^-}$ collision energies.
If the tau momentum directions are known one can construct
observables with a substantially higher sensitivity.
For the channels with only
two neutrinos in the final state the tau direction of
flight can be reconstructed up to a two-fold ambiguity.
This ambiguity can in principle be resolved by means of
the information obtained from a precise vertex detector
\cite{Kuhn}. Resolution of this ambiguity is, however,
not absolutely necessary (for details, see
\cite{O2,A3}).

One can  read off from the $\tau^+\tau^-$ production and decay
density matrices
the following CP- and T-odd observables for tracing
non-zero  $\rm{\mbox{Re}d^{\gamma,Z}_{\tau}}$.
(Below $\p,\m$ denote the momentum directions of the
charged final state particles taken  in the respective $\tau^+$ and $\tau^-$
rest systems).

\begin{eqnarray}
{\cal O}_1^{\rm Re} &=&
T^{\rm Re} + (\k\cdot\p)(\k\times\m)\cdot\e - (\k\cdot\m) (\k\times\p)\cdot\e ,
\nonumber\\
{\cal O}_2^{\rm Re} &=&
T^{\rm Re} + 4 (\e\times\k)\cdot(\p+\m) ,
\label{ReT}
\end{eqnarray}

\noindent where

\begin{equation}
T^{\rm Re} = - (\k\cdot\e)^2 (\p\times\m)\cdot\k +
(\k\cdot\e)(\p\times\m)\cdot\e .
\end{equation}

\noindent
The following CP- and CPT-odd observables are sensitive to
$\mbox{Im}d^{\gamma,Z}_{\tau}$:

\begin{eqnarray}
{\cal O}_1^{\rm Im} &=&
T^{\rm Im} + (\p+\m)\cdot\e - (\k\cdot\e) (\p+\m)\cdot\k , \nonumber \\
{\cal O}_2^{\rm Im} &=&
T^{\rm Im} + 4 (\k\cdot\p) (\e\cdot\m) - 4 (\k\cdot\m) (\e\cdot\p) ,
\label{ImT}
\end{eqnarray}

\noindent where

\begin{equation}
T^{\rm Im} = - (\k\cdot\e)^2 (\p + \m)\cdot\k + (\k\cdot\e)(\p+\m)\cdot\e.
\end{equation}
It is worth recalling that non-zero $\rm{{\mbox Im} d_{\tau}^{\gamma,Z}(s)}$
do not neccessarily require a $new$ production threshold $s_{thr} < s$.
Therefore it makes sense  to measure  the correlations (\ref{ImT}) even if
no new threshold has been discovered.

CP violation in tau decay would not leave its mark in the correlations
(\ref{ReT}),
(\ref{ImT}). For efficient CP tests in tau decay large
 samples of highly polarized $\tau^+$ and $\tau^-$ leptons are needed.

We have computed the expectation values of (\ref{ReT}),
(\ref{ImT})  in terms of the form factors for the above-mentioned channels
at the LEP2 energy $\sqrt s = $ 175 GeV and at
 $\sqrt s =$ 500 GeV (an energy relevant
for a linear $\rm{e^+e^-}$ collider). With these calculations one can
estimate the 1 s. d.
statistical errors with which the form factors can be measured. The results
are given  for an assumed number of events in Tables 1 - 4. Moreover, these
tables
contain also the results obtainable with optimal observables \cite{AS,O2,A3}
that have maximized  signal-to-noise ratios.

As to LEP2, one obtains practically the same results as those given in
Tables 1,2
at a somewhat higher energy, e.g. $\sqrt s =$ 190 GeV. The expected event
numbers at
LEP2 are roughly those of Tables 1,2. This means that  the real part of the
tau EDM can be measured with an accuracy of about $2\times 10^{-17}e$ cm.
This would be
a new result -- no $direct$ measurement of comparable sensitivity is available from
LEP1.  The leptoquark models discussed above indicate that $|{\mbox Re}
d_{\tau}^{\gamma}(s \simeq
175 {\rm GeV})|$ may be about eight times larger than  $|{\mbox Re}
d_{\tau}^{Z}(s = m_Z)|$.
This may serve as an incentive to measure this form factor.

The sensitivity to the EDM and WDM
is expected to increase with increasing c. m.
energy because, schematically, $<{\cal O}> \sim \sqrt s d_{\tau}(s)/e$.
In view of the a priori unknown
functional dependence  on $s$ of the form factors the accuracy estimates
of Tables 3,4 for $\sqrt s =$ 500 GeV may be taken as indication what can be
achieved at a linear collider. The numbers show that an interesting level of
sensitivity can be reached.
The event numbers used for these estimates correspond to an integrated
luminosity
of about 20${\rm fb}^{-1}$, which  is presently discussed \cite{Zerwas}.

In conclusion, the expectation values of the above observables are of the order
$<{\cal O}>\sim \sqrt s d_{\tau}(s)/e$ $ = (\sqrt s/m_Z)\delta$. We have
shown that $\delta$ can
be of the order of a few$\times 10^{-3}$. At LEP the OPAL and ALEPH
experiments  have shown
that correlations of this type can be measured with an accuracy of  a few
per mill.
Therefore it is worthwhile to perform such CP tests also at LEP2 and at a future
high luminosity linear $\rm{e^+e^-}$ collider.

\section*{Acknowledgements}

We wish to thank O. Nachtmann, A. Stahl, N. Wermes, and M. Wunsch
for many discussions concerning CP tests with tau leptons.

\section*{Figure Caption}
\begin{description}
\item{Fig. 1} The real and imaginary parts of the tau dipole moments
due to leptoquarks (model I)
in units of ${\rm Im}({\tilde\lambda}_1^*\lambda_1)\times 10^{-18}$$e$ cm
for $m_{\chi_1}$ = 200 GeV. EDM (a) and  WDM (b).
 \end{description}

\newpage
\def\white{\vrule width 0pt height 16pt depth 8pt}

\begin{center}

\begin{tabular}{|c|c||c|c||c|c|}\hline
 \multicolumn{2}{|c||}{}
& \multicolumn{2}{c||}{${\rm Re\ } d_\tau^\gamma$ \enskip
 $[10^{-18}e \rm cm]$}
& \multicolumn{2}{c|}{${\rm Re\ } d_\tau^Z$ \enskip
 $[10^{-18}e \rm cm]$} \white \\ \hline
\white
Channel & Events & \quad${\cal O}_1^{\rm Re}\quad$ & Optimal
& \quad${\cal O}_2^{\rm Re}$\quad & Optimal  \\ \hline \hline
\white$\pi-\pi$   & 100 &  44 & 35 & 17 & 15 \\
\white$\pi-\rho$  & 400 &  48 & 43 & 11 & 10 \\
\white$\rho-\rho$ & 400 & 106 & 95 & 18 & 17 \\
\white$\ell-\ell$ & 800 & 142 & 61 & 17 & 11 \\
\white$\ell-\pi$  & 600 & 55  & 33 & 19 & 8 \\
\white$\ell-\rho$ &1200 & 85  & 53 & 73 & 9.8 \\
\hline \hline
\white combined   &     & 25  & 18 & 6.9&  4.4 \\
\hline
\end{tabular}

\bigskip

\noindent{Table 1: 1 s. d. accuracy with which the real parts of
the $\tau$ dipole form factors can be measured at $\sqrt s = 175$ GeV
for a given number of events.
The event numbers include the charge conjugated modes.}

\bigskip\bigskip

\begin{tabular}{|c|c||c|c||c|c|}\hline
 \multicolumn{2}{|c||}{}
& \multicolumn{2}{c||}{${\rm Im\ } d_\tau^\gamma$ \enskip
 $[10^{-18}e \rm cm]$}
& \multicolumn{2}{c|}{${\rm Im\ } d_\tau^Z$ \enskip
 $[10^{-18}e \rm cm]$} \white \\ \hline
\white
Channel & Events & \quad${\cal O}_1^{\rm Im}\quad$ & Optimal
& \quad${\cal O}_2^{\rm Im}$\quad & Optimal  \\ \hline \hline
\white$\pi-\pi$   & 100  & 24  & 20 & 29 & 23 \\
\white$\pi-\rho$  & 400  & 17  & 15 & 32 & 28 \\
\white$\rho-\rho$ & 400  & 28  & 26 & 71 & 60 \\
\white$\ell-\ell$ & 800  & 27  & 17 & 87 & 39 \\
\white$\ell-\pi$  & 600  &  32 & 12 & 34 & 21 \\
\white$\ell-\rho$ &1200  & 123 & 15 & 54 & 3 \\
\hline \hline
\white combined   &      &  11 &6.6 & 16 & 12 \\
\hline
\end{tabular}

\bigskip

\noindent{Table 2: 1 s. d. accuracy with which the imaginary parts of
the $\tau$ dipole form factors can be measured at $\sqrt s = 175$ GeV.}

\begin{center}

\begin{tabular}{|c|c||c|c||c|c|}\hline
 \multicolumn{2}{|c||}{}
& \multicolumn{2}{c||}{${\rm Re\ } d_\tau^\gamma$ \enskip
 $[10^{-18}e \rm cm]$}
& \multicolumn{2}{c|}{${\rm Re\ } d_\tau^Z$ \enskip
 $[10^{-18}e \rm cm]$} \white \\ \hline
\white
Channel & Events & \quad${\cal O}_1^{\rm Re}\quad$ & Optimal
& \quad${\cal O}_2^{\rm Re}$\quad & Optimal  \\ \hline \hline
\white$\pi-\pi$   & 100 &  17 & 13 & 7.5 &  6.6 \\
\white$\pi-\rho$  & 400 &  18 & 16 & 4.9 &  4.3 \\
\white$\rho-\rho$ & 400 &  40 & 35 & 7.7 &  7.3 \\
\white$\ell-\ell$ & 800 &  54 & 22 & 7.3 &  5.0 \\
\white$\ell-\pi$  & 600 &  21 & 12 & 8.1 &  3.5 \\
\white$\ell-\rho$ & 1200&  32 & 20 & 31  &  4.2 \\
\hline \hline
\white combined   &     & 9.6 & 6.7&  3.0&  1.9 \\
\hline
\end{tabular}

\bigskip

\noindent{Table 3: 1 s. d. accuracy with which the real parts of
the $\tau$ dipole form factors can be measured at $\sqrt s = 500$ GeV.}

\bigskip\bigskip

\begin{tabular}{|c|c||c|c||c|c|}\hline
 \multicolumn{2}{|c||}{}
& \multicolumn{2}{c||}{${\rm Im\ } d_\tau^\gamma$ \enskip
 $[10^{-18}e \rm cm]$}
& \multicolumn{2}{c|}{${\rm Im\ } d_\tau^Z$ \enskip
 $[10^{-18}e \rm cm]$} \white \\ \hline
\white
Channel & Events & \quad${\cal O}_1^{\rm Im}\quad$ & Optimal
& \quad${\cal O}_2^{\rm Im}$\quad & Optimal       \\ \hline \hline
\white$\pi-\pi$   & 100 &  9.1  & 7.1 & 12 &  9.9 \\
\white$\pi-\rho$  & 400 &  6.5  & 5.3 & 14 & 12   \\
\white$\rho-\rho$ & 400 & 10    & 9.2 & 30 & 26   \\
\white$\ell-\ell$ & 800 &  10   & 6.2 & 37 & 17   \\
\white$\ell-\pi$  & 600 & 12    & 4.2 & 15 &  9.2 \\
\white$\ell-\rho$ &1200 &  47   & 5.2 & 23 & 15   \\
\hline \hline
\white combined   &     &  4.0  & 2.3 & 7.0&  5.1 \\
\hline
\end{tabular}

\bigskip

\noindent{Table 4: 1 s. d. accuracy with which the imaginary parts of
the $\tau$ dipole form factors can be measured at $\sqrt s = 500$ GeV.}

\end{center}

\end{center}

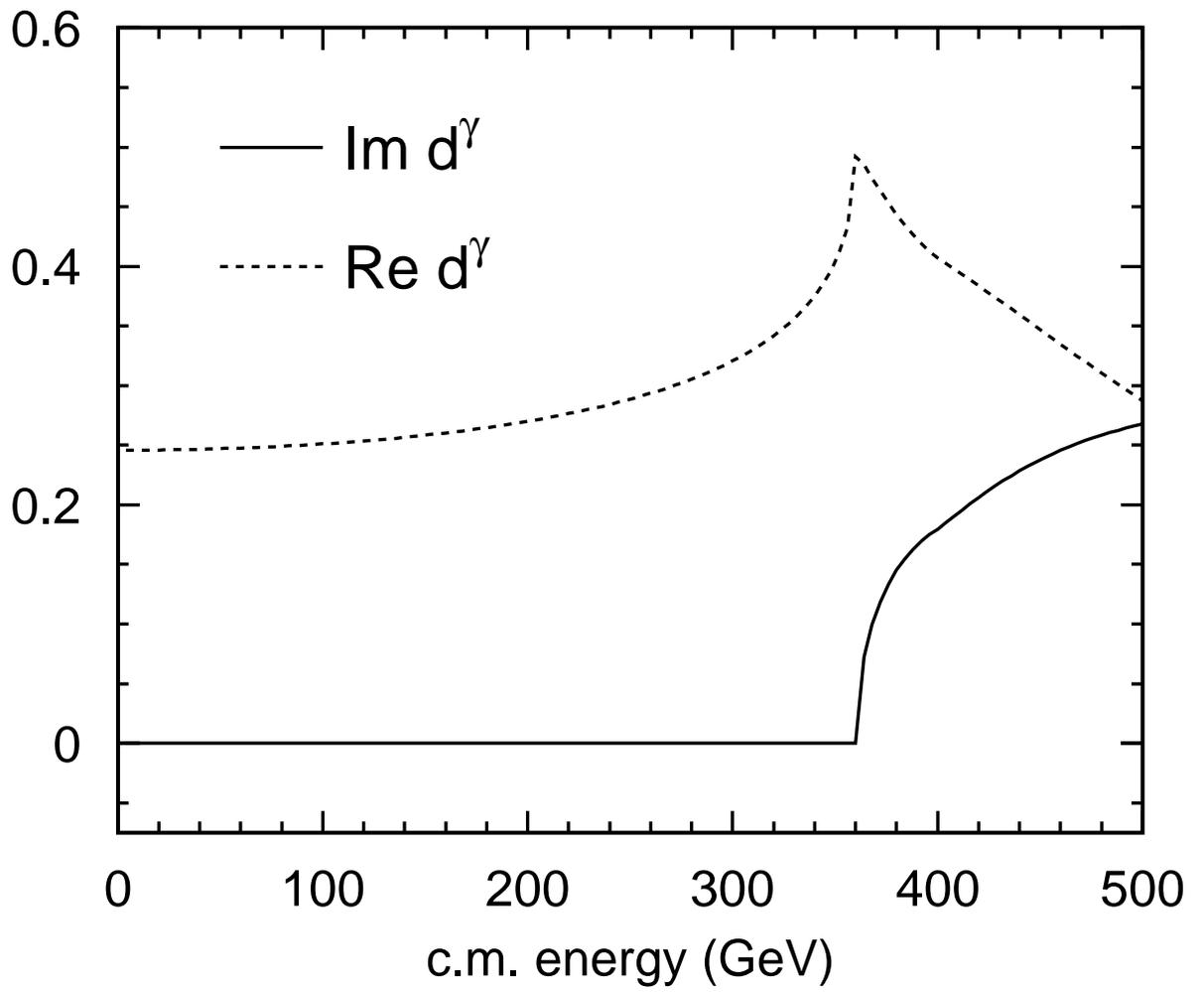
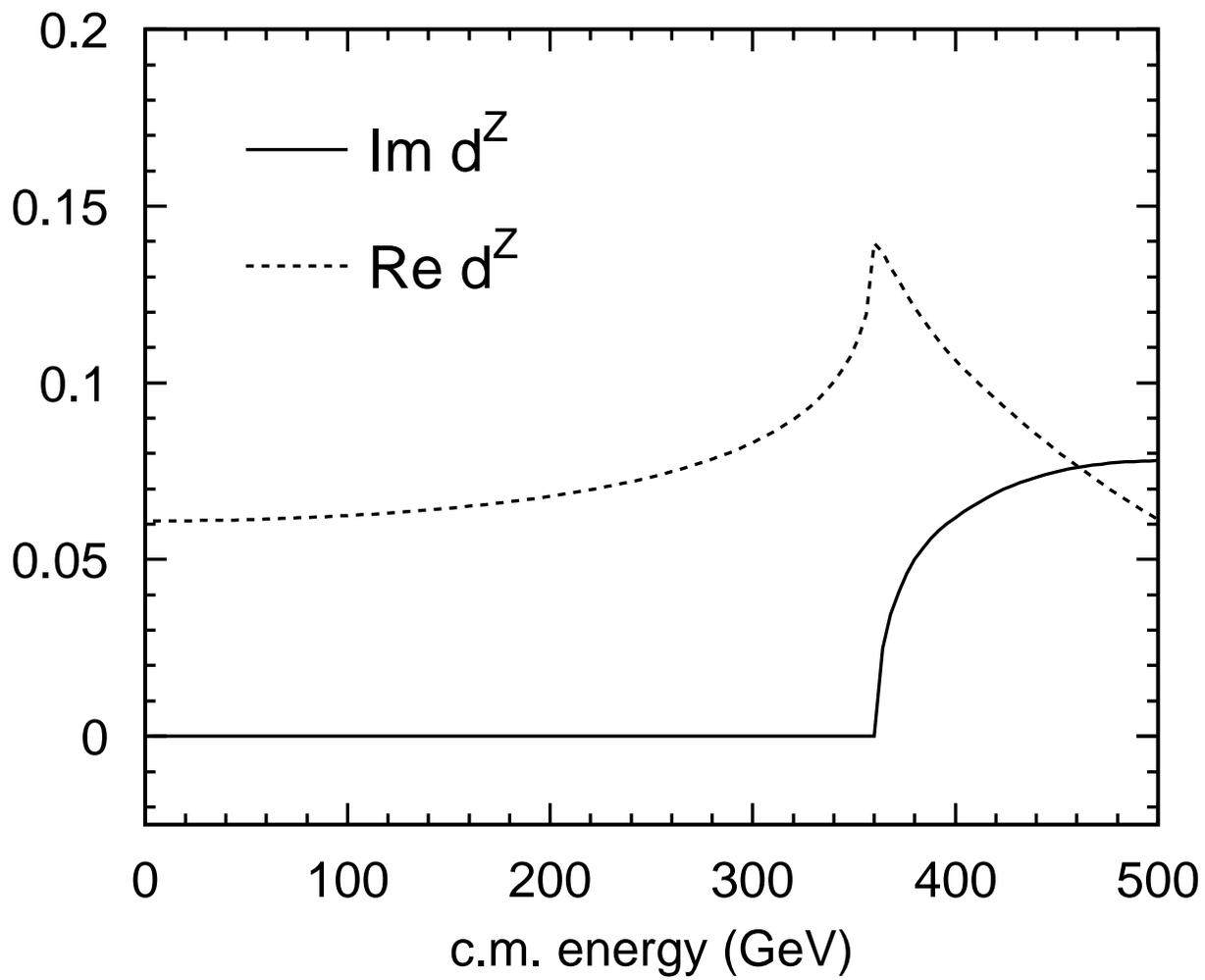
\end{document}